\documentclass[aip,jcp,reprint]{revtex4-1}

\usepackage{amsmath}
\usepackage{amsfonts}
\usepackage{url}
\usepackage{bm}
\usepackage{bbold}
\usepackage{graphics}
\usepackage{paralist}
\usepackage{graphicx}
\usepackage{afterpage}

\newcommand{\bra}{\langle}
\newcommand{\ket}{\rangle}
\newcommand{\bs}[1]{\boldsymbol{#1}}

\begin{document}

\title{Charge-transfer excited states: Seeking a balanced and efficient wave function ansatz in variational Monte Carlo}

\author{N. S. Blunt}
\email{nicksblunt@gmail.com}
\affiliation{Department of Chemistry, University of California, Berkeley, California 94720, USA}
\affiliation{Chemical Sciences Division, Lawrence Berkeley National Laboratory, Berkeley, California 94720, USA}
\author{Eric Neuscamman}
\email{eneuscamman@berkeley.edu}
\affiliation{Department of Chemistry, University of California, Berkeley, California 94720, USA}
\affiliation{Chemical Sciences Division, Lawrence Berkeley National Laboratory, Berkeley, California 94720, USA}

\begin{abstract}
We present a simple and efficient wave function ansatz for the treatment of excited charge-transfer states in real-space quantum Monte Carlo methods. Using the recently-introduced variation-after-response method [J. Chem. Phys.~\textbf{145}, 081103 (2016)], this ansatz allows a crucial orbital optimization step to be performed beyond a configuration interaction singles expansion, while only requiring calculation of two Slater determinant objects. We demonstrate this ansatz for the illustrative example of the stretched LiF molecule, for a range of excited states of formaldehyde, and finally for the more challenging ethylene-tetrafluoroethylene molecule.
\end{abstract}

\maketitle

\section{Introduction}
\label{sec:intro}

Electronically excited molecules, including those with significant charge-transfer (CT) character, are of great importance in fields such as photochemistry and in many areas beyond. Examples include metal-to-ligand CT in coordination compounds and dyes,\cite{Prier2013, Vogler2000} photocatalysis in energy materials,\cite{Park2016, Bokarev2015} and CT-mediated singlet fission.\cite{Chan2013, Zeng2014, Beljonne2013} Despite the need to study such states in order to understand many important phenomena, excited state computational methods are typically less developed than their ground-state counterparts.

Time-dependent density functional theory (TDDFT)\cite{Runge1984} is perhaps the most commonly used method for excited states, and often provides good accuracy for valence excitations at a manageable computational cost. However, it notoriously underestimates the energy of CT states due to a lack of long-range exchange.\cite{Dreuw2003} The use of range-separated hybrid density functionals often greatly ameliorates this error,\cite{Dreuw2004, Isborn2013} but the problem remains challenging in many cases. Equation-of-motion coupled cluster (EOM-CC)\cite{Emrich1981, Stanton1993, Krylov2008} theory typically provides much more reliable results, accurately including dynamic correlation and the correct asymptotic $1/r$ energy dependence for CT states. However, the $\mathcal{O}(N^6)$ scaling of canonical EOM-CCSD (where $N$ is a measure of system size) makes the theory too expensive for very large molecules.

Real-space quantum Monte Carlo methods,\cite{Foulkes2001, Needs2009} perhaps most notably variational Monte Carlo (VMC) and diffusion Monte Carlo (DMC),\cite{Grimm1971, Anderson1975, Umrigar1993} are important methods in electronic structure theory, with DMC often providing chemical accuracy for a reasonably chosen trial wave function. While the stochastic nature of such methods leads to a large prefactor in the computational cost, they have excellent parallel efficiency and relatively low scaling with system size, usually $\mathcal{O}(N^4)$ (or $\mathcal{O}(N^3)$ per sample). This has allowed such quantum Monte Carlo (QMC) methods to provide among the most accurate results available for systems of significant sizes, including solids.\cite{Foulkes2001, Kolorenc2011, Binnie2010}

However, QMC has historically been primarily applied to the study of ground states. Applications of QMC to excited states, both in discrete\cite{Purwanto2009, Booth2012_3, Blunt2015_3, Ten-no2013, Zhao2016_2} and real-space approaches,\cite{Ceperley1988, Caffarel1992, Williamson1998, Schautz2004_2, Filippi2009} have been more limited. In VMC, targeting excited states is challenging in large part due to the lack of appropriate target functions to minimize, although variance minimization, state-averaged minimization of energy, and other approaches have been used successfully, as in the work of Filippi and coworkers.\cite{Schautz2004_2, Filippi2009} The application of DMC to excited states\cite{Williamson1998, Schautz2004_2, Filippi2009} is challenging due to the lack of accurate and efficient trial wave functions, with a need to optimize trial states to great accuracy first, often by VMC.

Schautz, Buda and Filippi\cite{Schautz2004} considered the calculation of excitation energies with DMC in small photoactive molecules, using basic wave functions consisting of a Jastrow factor and a small determinantal expansion. It was found that optimization of the wave function, with respect to both orbitals and expansion coefficients, could drastically improve the quality of subsequent DMC calculations. Without this optimization, DMC excitation energies for formaldimine could be in error by more than $1$-$2$eV.

In this article we build on recent developments to optimize an efficient-to-calculate wave function ansatz in VMC, with a particular emphasis on CT states. This ansatz follows the recently-introduced variation-after-response (VAR) approach.\cite{Neuscamman2016} Specifically, we consider the linear response space of an optimizable determinant, which contains the full flexibility of a configuration interaction singles (CIS) wave function,\cite{Dreuw2005} together with orbital rotations. Using a finite-difference approximation, this wave function may be expressed efficiently as a difference of two Slater-Jastrow functions. By optimizing orbitals separately for each excited state, significant improvements are obtained for CT states compared to CIS, capable of correcting excitation energies by multiple electronvolts. Unlike the original VAR presentation,\cite{Neuscamman2016} the approach presented in this article is performed in real space, and therefore has a lower polynomial scaling of $\mathcal{O}(N^4)$, while also avoiding the costly construction and use of two-electron integrals.

In Section~\ref{sec:cis_bias} we discuss the error in excitation energies from the CIS method. In Section~\ref{sec:theory} we introduce our wave function form and discuss the underlying theory and scaling of the ansatz in VMC. Results are presented in Section~\ref{sec:results}. The stretched LiF molecule is considered as a simple example with clear low-lying CT states, including fixed-node diffusion Monte Carlo results. Formaldehyde is then considered, applying our ansatz to the nine lowest singlet excited states, demonstrating a range of states in a more typical molecule. Finally, a CT state is studied for the ethylene-tetrafluoroethylene (ETFE) system.

\section{Configuration interaction singles bias}
\label{sec:cis_bias}

Configuration interaction singles (CIS) is perhaps the most basic method for treating excited states. The CIS wave function is formed in the space of all single excitations from the Hartree--Fock (HF) determinant, $|D_0\ket$,
\begin{equation}
| \Psi_{\textrm{CIS}} \ket = \sum_{ai} \mu_{ai} \hat{a}_{a}^{\dagger} \hat{a}_i |D_0\ket,
\end{equation}
where $i$ and $a$ label occupied and virtual orbitals in the canonical HF basis, respectively. This wave function form allows a basic description of states involving excitation of only a single electron. The CIS method is computationally cheap, variational, and also gives the correct $1/r$ dependence\cite{Dreuw2003} on charge separation, $r$, for CT states. However, it accounts for no dynamic correlation and relies on orbitals that are biased in favor of the ground state, and as a result typically yields excitation energies which are too large, sometimes substantially so.

Subotnik recently investigated this error,\cite{Subotnik2011} demonstrating clearly that the bias is larger for CT states than for non-CT states, in some cases by more than $2$eV. The reason for this is clear: the HF orbitals, from which the CIS excited states are constructed, are optimized for the ground-state wave function. For non-CT states, each region of the molecule will have roughly the same overall electron density, and the same orbitals will allow for a qualitatively correct description of the state. In CT states, however, electrons transfer between regions of the molecule, with orbitals needing to relax for the new charge distribution. The basic CIS ansatz does not allow such orbital re-optimization, although effective relaxation can occur indirectly through the perturbative inclusion of double excitations, as in the study of Subotnik,\cite{Subotnik2011} and as originally considered by Head-Gordon \emph{et al.}\cite{Head-Gordon1994} EOM-CCSD also allows this relaxation indirectly through the inclusion of double excitations, which simultaneously account for dynamic correlation.

Liu \emph{et al.} considered a different approach where the CIS ansatz was applied together with a single orbital optimization step, similar to a step of the Newton-Raphson algorithm.\cite{Liu2012} This rotation step was found to significantly reduce bias in CT states relatives to non-CT states. While this optimization is not sufficient to achieve quantitative accuracy, it demonstrates how the qualitative description can be substantially corrected, and motivates additional development, as will be considered here.

\section{Theory}
\label{sec:theory}

\subsection{Optimizable determinants}
\label{sec:opt_dets}

To begin, consider a standard Slater determinant wave function,
\begin{equation}
\Psi_{\textrm{det}}(\bs{R}) = D^{\uparrow}(\bs{r}_1^{\uparrow}, \ldots, \bs{r}_{\frac{N}{2}}^{\uparrow}) D^{\downarrow}(\bs{r}_1^{\downarrow}, \ldots, \bs{r}_{\frac{N}{2}}^{\downarrow}),
\label{eq:slater_prod}
\end{equation}
where a separate Slater determinant is used for spin-up and spin-down electrons, as is standard in real-space QMC. We assume throughout this article that there are an equal number of spin-up and spin-down electrons, $N/2$ of each, and only consider restricted HF basis sets. $\bs{R}$ collectively denotes all electron positions.

We consider a single-particle basis of $M$ (spatial) molecular orbitals, $\{ \phi_1, \ldots, \phi_M \}$. The first $\frac{N}{2}$ of these are occupied in the HF determinant, leaving $M - \frac{N}{2}$ virtual orbitals. The molecular orbitals are themselves a linear combination of atomic orbitals, $\{ \chi_{\mu} \}$,
\begin{equation}
\phi_p(\bs{r}) = \sum_{\mu} \chi_{\mu} (\bs{r}) C_{\mu p}.
\end{equation}
Orbital rotations are then introduced via
\begin{equation}
\bs{C} = \bs{C}^{0} \bs{U},
\label{eq:orb_opt}
\end{equation}
with
\begin{equation}
\bs{U} = e^{-\bs{X}},
\label{eq:orb_opt_param}
\end{equation}
where $\bs{X}$ is an anti-symmetric matrix. $\bs{C}^{0}$ denotes the initial, Hartree--Fock, coefficient matrix. Thus, orbital optimization is parameterized by elements $X_{pq} = - X_{qp}$. To avoid redundancies, only rotations between occupied and virtual orbitals are allowed.

Each of $D^{\uparrow}$ and $D^{\downarrow}$ is formed as
\begin{align}
D &= |\phi_1 \phi_2 \ldots \phi_{\frac{N}{2}}|, \\
  &= \textrm{det}(\bs{A}),
\label{eq:det_def}
\end{align}
with
\begin{equation}
A_{ij} = \phi_j(\bs{r}_i)
\end{equation}
being the $\frac{N}{2} \times \frac{N}{2}$ Slater matrix.

Thus, the form used for the determinantal part of the wave function is given by Eq.~(\ref{eq:slater_prod}), with $D^{\uparrow / \downarrow}$ determined by Eq.~(\ref{eq:det_def}) and orbitals optimized via Eqs.~(\ref{eq:orb_opt}) and~(\ref{eq:orb_opt_param}).

\subsection{The finite-difference linear response (FDLR) wave function}
\label{sec:fdlr_wfn}

The above wave function form is appropriate for a single-reference ground state. As discussed in Section~\ref{sec:cis_bias}, CIS is a more natural starting point for (single-excitation) excited states. Such an expansion could be formed directly, and the expansion coefficients treated as optimizable parameters. Such expansions can be performed very efficiently, and offer great promise for studying excited states.\cite{Clark2011, Filippi2016, Assaraf2017}

As an alternative approach, we here make use of the fact that the CIS space is the linear response (LR) space of a determinant with orbital rotations, as presented in Sec.~(\ref{sec:opt_dets}). This is most clear by working in second quantization, where an optimizable determinant (in a restricted basis) may be expressed as
\begin{equation}
| D (\bs{X}) \ket = \textrm{exp} ( - \sum_{p > q} X_{pq} \hat{E}^{-}_{pq} ) \: | D_0 \ket,
\label{eq:opt_det_second_quant}
\end{equation}
where $p$ and $q$ run over spatial orbital labels, and with $\hat{E}^{-}_{pq} = \hat{E}_{pq} - \hat{E}_{qp}$ and $\hat{E}_{pq} = \hat{a}_{p \uparrow}^{\dagger} \hat{a}_{q \uparrow} + \hat{a}_{p \downarrow}^{\dagger} \hat{a}_{q \downarrow}$.
The LR space is spanned by first derivatives of the wave function with respect to its parameters. In this case, for a single determinant $| D (\bs{X}) \ket$, which only has orbital rotation parameters, $\{ X_{pq} \}$, the LR wave function is
\begin{align}
| \Psi_{\textrm{LR}} (\bs{\mu}, \bs{X} ) \ket &= \sum_{pq} \mu_{pq} \: \frac{\partial | D (\bs{X}) \ket }{\partial X_{pq}}, \\
                                     &= \sum_{pq} \mu_{pq} \: \hat{E}^{-}_{pq} | D_0 \ket,
\end{align}
which has the freedom of a general CIS wave function (with even-$S$ quantum number, due to enforcing $X^{\uparrow}_{pq} = X^{\downarrow}_{pq}$ in Eq.~(\ref{eq:opt_det_second_quant}), although it is simple to study the odd-$S$ space by enforcing $X^{\uparrow}_{pq} = -X^{\downarrow}_{pq}$). For the present application to real-space QMC, we work with the real-space determinant expressions from Sec.~(\ref{sec:opt_dets}), but the LR idea still applies.

This perhaps suggests working directly with
\begin{align}
\Psi_{\textrm{LR}} (\bs{\mu}, \bs{X}) &= \sum_{pq} \mu_{pq} \: \frac{\partial \Psi_{\textrm{det}} (\bs{X}) }{\partial X_{pq}}, \\
                                                  &= \sum_{pq} \mu_{pq} \: \frac{\partial \big[ D^{\uparrow} (\bs{X}) D^{\downarrow} (\bs{X}) \big] }{\partial X_{pq}},
\label{eq:exact_der}
\end{align}
since determinant derivatives are easily derivable and efficient to calculate. However, VMC optimization of a wave function (by, for example, the linear method\cite{Umrigar2007}) requires evaluation of first-order parameter derivatives, $\partial \Psi_{\textrm{LR}} / \partial X_{pq}$. This then requires second-order parameter derivatives of $\Psi_{\textrm{det}}$, which are inefficient to calculate compared to existing alternative QMC and electronic structure methods.

We therefore instead consider the following finite-difference approximation\cite{Neuscamman2016} to Eq.~(\ref{eq:exact_der}),
\begin{align}
\Psi_{\textrm{FDLR}} (\bs{\mu}, \bs{X}) &= \Psi_{\textrm{det}} (\bs{X} + \bs{\mu}) - \Psi_{\textrm{det}} (\bs{X} - \bs{\mu}), \\
                                                    &\equiv \Psi_{\textrm{det}}^+ - \Psi_{\textrm{det}}^-,
\end{align}
where the second line defines our shorthand for both terms. We refer to this as the finite difference linear response (FDLR) wave function. In the limit of small $\bs{\mu}$ parameters, this ansatz is exactly equivalent to Eq.~(\ref{eq:exact_der}), up to  a normalization factor. However, use of this ansatz only requires calculation of two $\Psi_{\textrm{det}}$ objects. Previous studies have been performed on linear response in VMC.\cite{Zhao2016_2, Mussard2017} In the present study, we consider not only linear response around the ground-state ansatz, but allow re-optimization of this underlying wave function for each excited state: variation-after-response.\cite{Neuscamman2016} For the linear response of Hartree--Fock theory, this relaxation is simply equivalent to orbital optimization for each excited state. Setting $\bs{X}$ to $\bs{0}$ and optimizing with respect to $\bs{\mu}$ (in the absence of a Jastrow factor, and in the small $\bs{\mu}$ limit) is equivalent to a CIS calculation. Optimizing with respect to $\bs{X}$ will perform orbital optimization.

One may worry that random sampling in the presence of a finite-difference approximation will lead to large errors. However, both $\Psi_{\textrm{det}}^+$ and $\Psi_{\textrm{det}}^-$ components are calculated from identical samples, and so $\Psi_{\textrm{FDLR}}$ is calculated correctly per sample (although the finite-difference approximation itself remains). One may further worry that small $\bs{\mu}$ parameters may be difficult to optimize while simultaneously optimizing with respect to larger $\bs{X}$ and Jastrow parameters. Ultimately, this will be tested through application, although we do not find a very small $\bs{\mu}$ to be necessary for accurate results, as discussed in Section~\ref{sec:formaldehyde}. Furthermore, optimization of $\bs{X}$ parameters alone is often sufficient to dramatically improve results, as we shall see.

Finally, spline-based electron-nuclear and electron-electron Jastrow factors are included, denoted collectively as $J(\bs{R})$, so that the total wave function form is
\begin{align}
\Psi_{\textrm{tot}}(\bs{R}) &= \textrm{J}(\bs{R}) \: \Psi_{\textrm{FDLR}}(\bs{R}), \\
                            &= \textrm{J}(\bs{R}) \: \big[ \Psi_{\textrm{det}}^+(\bs{R}) - \Psi_{\textrm{det}}^-(\bs{R}) \big].
\label{eq:FDLR}
\end{align}
We seek to optimize this wave function form with respect to all parameters, thus achieving variation-after-response independently for each excited state.

Because it is formed as the difference of two Slater-Jastrow functions, VMC and DMC simulations using the FDLR ansatz will have the same scaling as traditional real-space QMC calculations. Per sample, this scaling consists of $\mathcal{O}(N^2)$ terms for the two-body Jastrow, electron-electron and electron-ion terms, and $\mathcal{O}(N^3)$ scaling for construction of the Slater matrix and evaluation of relative determinant values after electron moves.\cite{Williamson2001, Ahuja2011} Thus, the per-sample scaling of VMC and DMC is $\mathcal{O}(N^3)$. The number of samples required for a fixed error is typically $\mathcal{O}(N)$, for systems of up to roughly a few hundred electrons,\cite{Towler2009} and so our overall scaling is $\mathcal{O}(N^4)$. For comparison, EOM-CCSD and EOM-CCSDT scale as $\mathcal{O}(N^6)$ and $\mathcal{O}(N^8)$, respectively. We note that deterministic methods like CC theory are amenable to approaches such as density fitting and locality approximations, which can greatly reduce their scaling.\cite{Whitten1973, Werner2003, Neese2009, Schutz2013} However, locality approximations can also be used to reduce the dominant term in VMC and DMC simulations to linear scaling per sample.\cite{Williamson2001} The low-polynomial scaling of the FDLR ansatz should make this excited-state approach viable for systems containing hundreds of electrons. In Section~\ref{sec:etfe} of this initial presentation, ETFE is considered, treating 48 electrons using considerably lower computational resources than state-of-the-art QMC calculations.

To perform CIS with orbital rotations by traditional approaches, the cost would be $\mathcal{O}(N^5)$. This is due to the cost of transforming two-electron integrals to the rotated basis. Although, as discussed above, reduced-scaling approaches could be used to improve this situation (and once again we note that such approaches can be used in real-space QMC). However, we emphasize that the advantage of our approach over CIS with orbital rotations is far greater than improved scaling. Because we work in the framework of real-space QMC, we can trivially apply Jastrow factors to include a substantial proportion of dynamic correlation, entirely absent from basic CIS with orbital rotations. Furthermore, these optimized trial wave functions can be used in fixed-node DMC, which has the same scaling as VMC (allowing large systems to be treated), but is often capable of chemical accuracy at the basis set limit.

\subsection{Excited-state variational principle}
\label{sec:var_principle}

We use a recently-introduced variational principle\cite{Zhao2016} for direct targeting of excited states. Consider the following target function, $\Omega$:
\begin{align}
\Omega(\Psi, \omega) &= \frac{ \bra \Psi | (\omega - \hat{H} ) | \Psi \ket }{ \bra \Psi | (\omega - \hat{H} )^2 | \Psi \ket }, \\
                     &= \frac{ \omega - E }{ (\omega - E)^2 + \sigma^2 },
\label{eq:target_fn}
\end{align}
where $E = \bra \Psi | \hat{H} | \Psi \ket / \bra \Psi | \Psi \ket$ is the local energy and
\begin{equation}
\sigma^2 = \frac{ \bra \Psi | (\hat{H} - E)^2 | \Psi \ket }{ \bra \Psi | \Psi \ket }
\end{equation}
is the variance.

The shift $\omega$ is chosen to allow specific states to be targeted by the optimizer. Zhao and Neuscamman proved that the global minimum of $\Omega(\Psi)$ (for a fixed $\omega$) is achieved by setting $| \Psi \ket$ to equal the eigenstate of $\hat{H}$ whose eigenvalue lies directly above $\omega$. Thus, if one can optimize $\Omega$ (for a fixed $\omega)$ to its global minimum for an arbitrarily flexible ansatz, then a specific eigenstate is guaranteed to have been reached. This is in contrast to the variance minimization approach to excited states, in which one makes use of the fact that any exact eigenstate of $\hat{H}$ has zero variance. A downside to this approach is that each eigenstate yields the same optimal value of $\sigma^2=0$, and so the state reached depends only on the parameters initially guessed.

In practice, exact eigenstates of $\hat{H}$ will not lie in the parameter space for the ansatz considered. As a result, it is more difficult to make rigorous statements about which state will be the global minimum of $\Omega(\omega)$ for a given value of $\omega$. However, we note that for a fixed $\Psi$, $\Omega(\omega)$ is minimized by setting $\omega = E - \sigma$, as can be seen by investigating Eq.~(\ref{eq:target_fn}). This, together with sensible initial guesses, has allowed us to optimize the desired excited states in this article without substantial difficulty, and without ambiguity in the choice of $\omega$.

For the FDLR ansatz of Eq.~(\ref{eq:FDLR}), one can always set $\bs{X}=\bs{0}$ and take CIS coefficients (appropriately scaled) as initial parameters for $\bs{\mu}$, in order to start the optimization from the HF-basis CIS solution. This CIS calculation will always be cheaper than the following VMC optimization, and so this is a sensible initial guess. From here, we perform a VMC calculation such that $E - \sigma$ can be calculated with greater accuracy than the gaps between energy eigenvalues. This then determines both $\omega$ and $\Psi$ from which to perform optimization. Currently, after a single optimization of $\Omega$, we restart the optimization with the latest value of $\omega = E - \sigma$. In practice we do not find that the state targeted changes, but the final energy may be slightly altered due to the more accurate $\omega$ value, typically by no more than a few m$E_{\textrm{h}}$, with energy differences varying even less.

The optimization of $\Omega$ is performed using a modified version of the linear method, as described in Ref.~(\onlinecite{Zhao2016}).

\section{Results}
\label{sec:results}

\subsection{LiF}
\label{sec:LiF}

As a clear demonstration of optimization of the FDLR wave function for a simple CT state, the LiF molecule is considered at a stretched nuclear distance of $3.5$\AA.

The RHF ground state for this molecule is the ionic $\textrm{Li}^+ \textrm{F}^-$ solution, and as such the orbitals will be optimized for this charge distribution. However, at this stretched geometry there clearly exist low-lying neutral excited states, corresponding to both the lithium and fluorine atoms in their neutral ground state. At infinite separation, and considering only singlet states, this will be triply-degenerate: relative to the ionic ground state, one can consider exciting an electron from either $2p_x$, $2p_y$ or $2p_z$ orbitals on the fluorine to a $2s$ orbital on the lithium. For finite-separation, and defining the $z$-axis as the internuclear axis, the excitation from the $2p_z$ orbital will be slightly non-degenerate with the other two states. The excitations from $2p_x$ and $2p_y$ orbitals will be degenerate at any separation due to symmetry about the $z$-axis.

Here, the ground state is ionic and the excited states neutral, perhaps the opposite of the more typical CT situation. However, this nonetheless serves as a clear demonstration of the principle: the ground state orbitals will be inappropriate to describe the neutral states, and optimization of the FDLR wave function should lead to a substantial reduction in the excitation gap relative to CIS.

We use the pseudopotentials of Burkatzki \emph{et al.}\cite{Burkatzki2007} for both Li and F atoms, replacing 2 electrons from both, and the corresponding valence-double zeta (VDZ) basis set. While this may seem unnecessary for such a small system, we use this only as a demonstrative example, and moreover it leads to the interesting situation where the ground state has no electrons on the Li center. This perhaps suggests that there is little to constrain the HF orbitals on the lithium, and perhaps an even greater need for orbital optimization in the neutral state.

\begin{figure}[t!]
\includegraphics{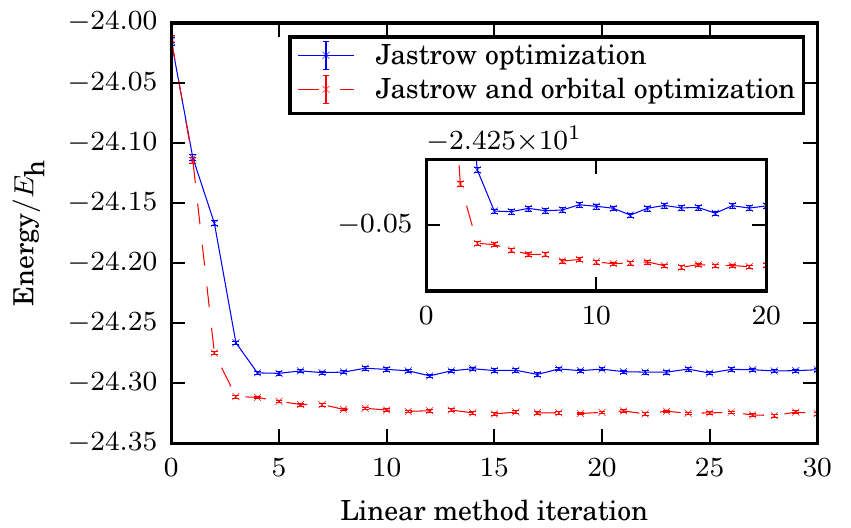}
\caption{Linear method convergence for the first excited state ($2p_x$ on F to $2s$ on Li transition) of LiF at a stretched geometry of $3.5$\AA. Pseudopotentials\cite{Burkatzki2007} and a corresponding VDZ basis are used. Convergence is shown for optimization of the Jastrow alone, and also for optimization of the Jastrow and orbitals simultaneously. Convergence takes around $5$ iterations in the former case compared to $10$-$15$ with orbital optimization, but the final energy is reduced by a further $\sim 32$m$E_{\textrm{h}}$.}
\label{fig:LiF_opt}
\end{figure}

\begin{table*}[t!]
\begin{center}
{\footnotesize
\begin{tabular}{@{\extracolsep{4pt}}lccccccccc@{}}
\hline
\hline
& \multicolumn{5}{c}{Excitation energy, $\Delta E$/$E_{\textrm{h}}$} & \multicolumn{4}{c}{Standard deviation, $\sigma$/$E_{\textrm{h}}$} \\
\cline{2-6} \cline{7-10}
&  &  & \multicolumn{3}{c}{VMC} & & \multicolumn{3}{c}{VMC} \\
\cline{4-6} \cline{8-10}
State ( $ \textrm{F} \rightarrow \textrm{Li}$ ) & CIS & EOM-CCSD & $\{ J \}$ & $\{ J, X \}$ & $\{ J, X, \mu \}$ & CIS & $\{ J \}$ & $\{ J, X \}$ & $\{ J, X, \mu \}$ \\
\hline
$ 1 \; (2p_x \rightarrow 2s) $ & 0.096 & 0.039 & 0.077 & 0.045 & 0.042 & 1.95 & 0.733(2)  & 0.701(1)  & 0.6977(4) \\
$ 2 \; (2p_y \rightarrow 2s) $ & 0.096 & 0.039 & 0.077 & 0.044 & 0.042 & 1.95 & 0.7306(8) & 0.6968(4) & 0.6959(4) \\
$ 3 \; (2p_z \rightarrow 2s) $ & 0.102 & 0.052 & 0.087 & 0.057 & 0.051 & 1.95 & 0.7316(5) & 0.708(3)  & 0.6982(9) \\
\hline
\hline
\end{tabular}
}
\caption{Excitation energies (from the ground state) and standard deviations for the first three excited states of LiF at a stretched geometry of $3.5$\AA. Statistical uncertainties where not given are sub-milli-Hartree. Pseudopotentials\cite{Burkatzki2007} and a corresponding VDZ basis are used. These states all occur via charge transfer relative to the ground state. For VMC calculations, parameters $\{ \ldots \}$ in curly brackets specify which parameters are optimized simultaneously by the linear method. CIS estimates of $\Delta E$ are too large by $\sim 50 - 60$m$E_{\textrm{h}}$, compared to EOM-CCSD benchmarks. In VMC, optimization of the Jastrow alone corrects the energy gap by $\sim 15 - 20$m$E_{\textrm{h}}$. Simultaneous optimization of the Jastrow and orbitals reduces $\Delta E$ to good agreement with EOM-CCSD. Further optimization of the CIS coefficients, $\mu$, leads to a further reduction of a few m$E_{\textrm{h}}$.}
\label{tab:LiF}
\end{center}
\end{table*}

We note that the ground state wave function, from which excitation energies are calculated, is always a single Jastrow-Slater wave function with orbital rotations. CIS wave functions are automatically orthogonal to the HF determinant, and so the FDLR ansatz would not be appropriate for the ground state. It is possible that the ground-state wave funcion may be improved in the presence of a Jastrow by additional determinants, but we expect any resulting change in energy to be very small.

Fig.~\ref{fig:LiF_opt} presents the convergence of the linear method, when optimizing the Jastrow factor alone, and when optimizing both the Jastrow and orbitals simultaneously, for the first excited state of LiF. This state corresponds to a $2p_x$ to $2s$ excitation from the fluorine to the lithium atom. The ground state has a dipole moment (from HF/CIS) of -3.196$e$\AA, while the first excited state has a dipole moment (from CIS) of -0.035$e$\AA, giving a relative dipole moment of 3.161$e$\AA, corresponding to almost a full transfer of charge $-e$, as expected. As can be seen, including orbital optimization reduces the energy by about 35m$E_\textrm{h}$ ($\sim 1$eV) compared to a Jastrow-only optimization. Convergence is somewhat slower with orbital optimization, but still takes only $10$-$15$ linear method iterations.

Table~\ref{tab:LiF} presents final excitation energies, compared to CIS and EOM-CCSD, and wave function standard deviation compared to CIS. We expect that the EOM-CCSD energies have good accuracy for this VDZ basis set, although we note that EOM-CCSD can also overestimate some excitation energies.\cite{Thiel2008} However, this overestimation should be very small compared to that of CIS, and EOM-CCSD is sufficient for the comparison here. A comparison with VMC results is non-trivial, because they contain a Jastrow factor which effectively makes them beyond-VDZ basis results. Larger bases can be used for EOM-CCSD results, but these then contain large amounts of dynamic correlation not present in the FDLR ansatz. For the purpose of demonstrating the effects of orbital optimization in CT states, the present comparison is sufficient. We will consider the application of diffusion Monte Carlo (DMC) using FDLR trial wave functions for LiF shortly, where a much higher fraction of correlation energy is captured.

The VMC optimization is performed starting from the CIS excited states (i.e., by setting $\bs{\mu}$ to the CIS coefficients in the FDLR ansatz, with $\bs{X} = \bs{0}$). CIS excitation energies are significantly overestimated, by $\sim 1.5$-$2$eV, compared to EOM-CCSD. Optimization of the Jastrow factor alone reduces the excitation energy by $\sim 15$-$20$m$E_{\textrm{h}}$, depending on the state. The fact that the Jastrow alone can reduce this gap is probably due to the simple nature of the system: in each excited state, there is only one electron on the lithium atom. The electron-nuclear Jastrow is therefore optimized entirely to account for this electron. Beyond enforcing the nuclear cusp condition, the electron-nuclear Jastrow effectively has the ability to either ``shrink'' or ``expand'' orbitals, as required for optimization. Since the lithium's electron is in a spherically-symmetric $2s$ orbital, the electron-nuclear Jastrow alone is able to qualitatively improve the wave function by itself. For larger systems, such as formaldehyde studied in Section~\ref{sec:formaldehyde} and ETFE studied in Section~\ref{sec:etfe}, optimization of the Jastrow has little effect on excitation energies.

Optimizing the Jastrow and orbitals simultaneously leads to a more substantial reduction, approximately $30$m$E_{\textrm{h}}$ beyond the Jastrow-only optimization, bringing results closely in line with EOM-CCSD values.

When optimizing all parameters together, $\{ J, X, \mu \}$, we choose to start the optimization from the result of a previous $\{ J, X \}$ optimization. We find that this makes the all-parameter optimization easier, and helps prevent optimization to the wrong state - a concern because all CIS excited states are reachable by varying $\mu$. By optimizing $\{ J, X \}$ parameters first, optimization of $\{ J, X, \mu \}$ begins from a lower point in the $\Omega$ landscape, where the optimization is more likely ``locked in'' to the correct state, and $\sigma$ is smaller.

The final $\{ J, X, \mu \}$ optimization gives a further energy reduction of a few m$E_{\textrm{h}}$. We typically find the relaxation of the CIS coefficients to give a smaller improvement than for the orbital rotation parameters, although some exceptions do occur, as will be seen in Sec.~(\ref{sec:formaldehyde}).

Importantly, the standard deviation of the wave function, $\sigma$, also decreases upon orbital optimization. We reiterate that we optimize $\Omega(\omega)$ rather than $\sigma^2$, but minimization of $\Omega$ should clearly reduce $\sigma$. The reduction in $\sigma$ is small compared to that from optimization of the Jastrow. However, unlike the Jastrow, optimization of orbitals leads to a correction in the wave function nodal surface, crucial for subsequent DMC simulations. Schautz \emph{et al.} previously found such optimization of orbitals and expansion coefficients to be necessary for accurate subsequent DMC excitation energies.\cite{Schautz2004}

\begin{table*}[t!]

\begin{center}
{\footnotesize
\begin{tabular}{@{\extracolsep{4pt}}lccccccccccc@{}}
\hline
\hline
& \multicolumn{7}{c}{Absolute energy, $(E+24E_{\textrm{h}})$/$E_{\textrm{h}}$} & \multicolumn{4}{c}{Excitation energy, $\Delta E$/$E_{\textrm{h}}$} \\
\cline{2-8} \cline{9-12}
&  \multicolumn{4}{c}{EOM-CCSD} & \multicolumn{3}{c}{DMC} & \multicolumn{1}{c}{EOM-CCSD} & \multicolumn{3}{c}{DMC} \\
\cline{2-5} \cline{6-8} \cline{9-9} \cline{10-12}
State & VDZ & VTZ & VQZ & V5Z & $\{ J \}$ & $\{ J, X \}$ & $\{ J, X, \mu \}$ & V5Z & $\{ J \}$ & $\{ J, X \}$ & $\{ J, X, \mu \}$ \\
\hline
Ground                         & -0.320 & -0.417 & -0.449 & -0.460 & -0.4638(8) & -0.4642(5) & - & - & - & - & - \\
$ 1 \; (2p_x \rightarrow 2s) $ & -0.281 & -0.348 & -0.368 & -0.374 & -0.3760(1) & -0.3772(1) & -0.3778(1) & 0.086 & 0.0878(8) & 0.0870(5) & 0.0864(5) \\
$ 2 \; (2p_y \rightarrow 2s) $ & -0.281 & -0.348 & -0.368 & -0.374 & -0.3760(1) & -0.3771(2) & -0.3778(1) & 0.086 & 0.0878(8) & 0.0871(5) & 0.0864(5) \\
$ 3 \; (2p_z \rightarrow 2s) $ & -0.268 & -0.341 & -0.362 & -0.369 & -0.3687(2) & -0.3708(2) & -0.3723(2) & 0.091 & 0.0951(8) & 0.0934(5) & 0.0919(5) \\
\hline
\hline
\end{tabular}
}
\caption{Absolute energies ($+24E_{\textrm{h}}$) and excitation energies (from the ground state) for the first three excited states of LiF at a stretched geometry of $3.5$\AA, using BFD pseudopotentials. Note that absolute energies are specific to this pseudopotential. Absolute energies demonstrate the extent of energy differences between basis sets, even for this simple system, and the extreme accuracy of DMC results. Optimization of orbitals by VMC reduces the subsequent DMC energy by $\sim 1-2$m$E_{\textrm{h}}$, depending on the state. Optimization of CIS coefficients leads to a similar reduction. For the ground state, optimization of orbitals results in a smaller energy change.}
\label{tab:LiF_dmc}
\end{center}
\end{table*}

\subsection{LiF (diffusion Monte Carlo)}

We now present fixed-node diffusion Monte Carlo energies for the same LiF system, for the same states considered above, and using optimized FDLR wave functions. The trial wave functions used are exactly those that were obatined from the final iteration of the above VMC calculations. Since pseudopotentials are in use, we use T-moves for all DMC simulations.\cite{Casula2006} All simulations used $1200$ walkers and a timestep of $0.001$.

In contrast to VMC with basic one- and two-body Jastrow factors, DMC is often capable of extreme accuracy, even in the basis set limit. For an accurate comparison, we require benchmarks that are in a very large basis set. We once again use equation-of-motion coupled cluster with single and double excitations (EOM-CCSD), but now consider basis sets up to quintuple-zeta.

Results are shown in Table~\ref{tab:LiF_dmc}, presenting both absolute energies and excitation energies relative to the ground state. Note that BFD pseudopotentials are applied, and absolute energies are therefore calculated with pseudopotentials in use. Absolute energies are shown against EOM-CCSD results from various basis sets. To obtain similar results to DMC, EOM-CCSD must use an extremely large quintuple-zeta basis set. Even from VQZ to V5Z, the ground state energy is reduced by a further 11m$E_{\textrm{h}}$. This demonstrates the great difficulty of obtaining accurate benchmarks against which DMC results can be compared. This is particularly true in excited states, where basis set errors tend to be particularly large, and where there are fewer high-accuracy methods available.

For the ground state, the energy from DMC is below that of V5Z EOM-CCSD by around 4m$E_{\textrm{h}}$. For excited states, fully-optimized trial wave functions ($ \{ J, X, \mu \} $) lead to absolute energies of around 3m$E_{\textrm{h}}$ lower than those of V5Z EOM-CCSD. For the ground state, for which the RHF orbitals used are appropriate, further orbital optimizaton of the trial wave function has no effect within error bars. For the excited states (which involve charge transfer, relative to the ground state), optimization of orbitals leads to a lowering of absolute energies by around $1-2$m$E_{\textrm{h}}$, for the states considered. Further optimization of CIS coefficients ($\mu$) leads to a similar reduction. Clearly, this is a small change, but this is also a very simple system. For larger and more complicated systems, we expect the effect of orbital optimization on CT states, and in general, to be more significant.\cite{Schautz2004} We do not consider DMC further in this article, but work on DMC using FDLR in larger and less trivial systems is underway.

\subsection{Formaldehyde}
\label{sec:formaldehyde}

Formaldehyde is a simple but important molecule in many chemical processes. Formaldehyde is a common product of combustion\cite{Garner1939} and an important photoactive biomolecule, while electronically excited formaldehyde specifically is known to be responsible for blue color of cool flames\cite{Sheinson1973} among other phenomena.

Formaldehyde is considered at its equilibrium, ground state geometry. The lowest nine excited states (as determined by CIS) of singlet symmetry are studied, considering all point group symmetry sectors. Pseudopotentials from the set of Burkatzki \emph{et al.}\cite{Burkatzki2007} are used for C and O atoms. HF and CIS calculations were performed with \url{GAMESS}.\cite{GAMESS} EOM-CCSD calculations were performed with \url{Molpro}.\cite{MOLPRO}

Formaldehyde presents a different situation to that of LiF. It is not an obvious candidate for states with strong CT character, and there are a variety of effects present among the nine excited states considered. This includes varying amounts of dynamic correlation, which the FDLR ansatz is not expected to capture fully. However, the application of orbital rotations and relaxation of CIS coefficients should nonetheless yield important improvements to the CIS wave functions.

\begin{figure}[t!]
\includegraphics{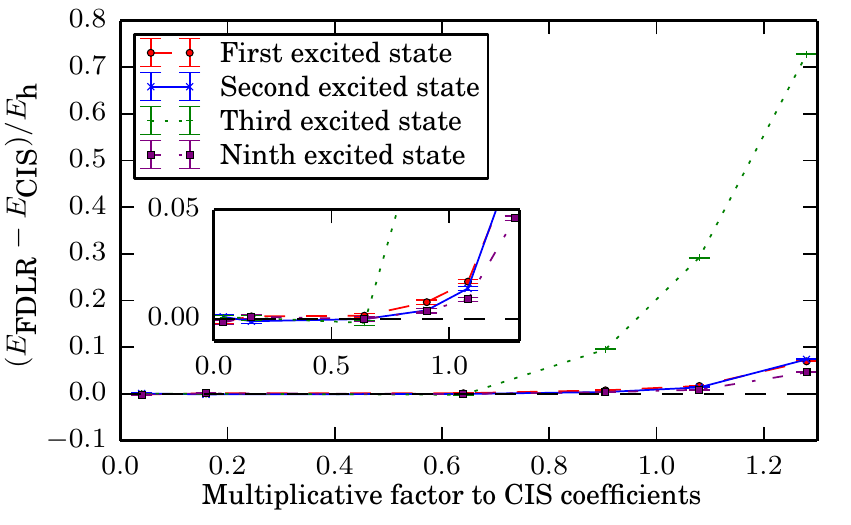}
\caption{FDLR energies (with no Jastrow factor and no rotations applied to the RHF orbitals) relative to CIS energies, as the magnitude of $\bs{\mu}$ parameters in the FDLR ansatz are varied. The $\bs{\mu}$ parameters are set equal to the CIS coefficients of the desired excited state, multiplied by some small number. The $x$-axis gives the value of this multiplicative factor. This factor should be small enough that the finite-difference approximation is sufficiently accurate. The lowest three and ninth singlet excited states are considered. The error in the third excited state grows most quickly as the magnitude of $\bs{\mu}$ is increased. However, it is seen that $\bs{\mu}$ parameters need not be very small to achieve sufficient accuracy: better than $1$m$E_{\textrm{h}}$ agreement with CIS occurs with a factor of $0.64$, for each state. The ninth excited state is no more challenging than the first or second excited states, suggesting that highly-excited states are not a particular challenge. We use a factor of $0.01$ for all results presented in this article.}
\label{fig:ch2o_converge}
\end{figure}
We first address a concern raised in Section~\ref{sec:theory}, namely the magnitude of $\bs{\mu}$ parameters required to accurately replicate CIS. The FDLR ansatz may only exactly replicate CIS in the limit $|\bs{\mu}| \to 0$. However, we may ask at what point milli-Hartree accuracy may be achieved. Fig.~(\ref{fig:ch2o_converge}) shows convergence of the FDLR energy (without a Jastrow factor and using RHF orbitals, $\bs{X}=\bs{0}$) for varying magnitudes of $\bs{\mu}$. To form the FDLR wave function, CIS coefficients are generated with an $L^2$ norm of $1$. These are then multiplied by a constant factor to give the $\bs{\mu}$ used in the intial, unoptimized FDLR wave function. The $x$-axis of Fig.~(\ref{fig:ch2o_converge}) denotes this multiplicative factor. The $y$-axis then plots the energy difference relative to CIS, for the first three excited states, and the ninth excited state. It is seen that accuracy within $1$m$E_{\textrm{h}}$ is achieved with a factor as large as $0.64$ in all cases. In practice, we have set this factor to $0.01$ in all calculations presented in this article, significantly smaller than the required value (although this will be system dependent). The ninth excited state requires a similar magnitude of $\bs{\mu}$ as the first and second excited states, and we do not find highly-excited states to be particularly challenging in this regard. We note that a constant normalization of $\bs{\mu}$ is not enforced, and the optimizer is free to reduce or increase this magnitude as required for minimization of the target function.

\begin{figure*}[t!]
\includegraphics{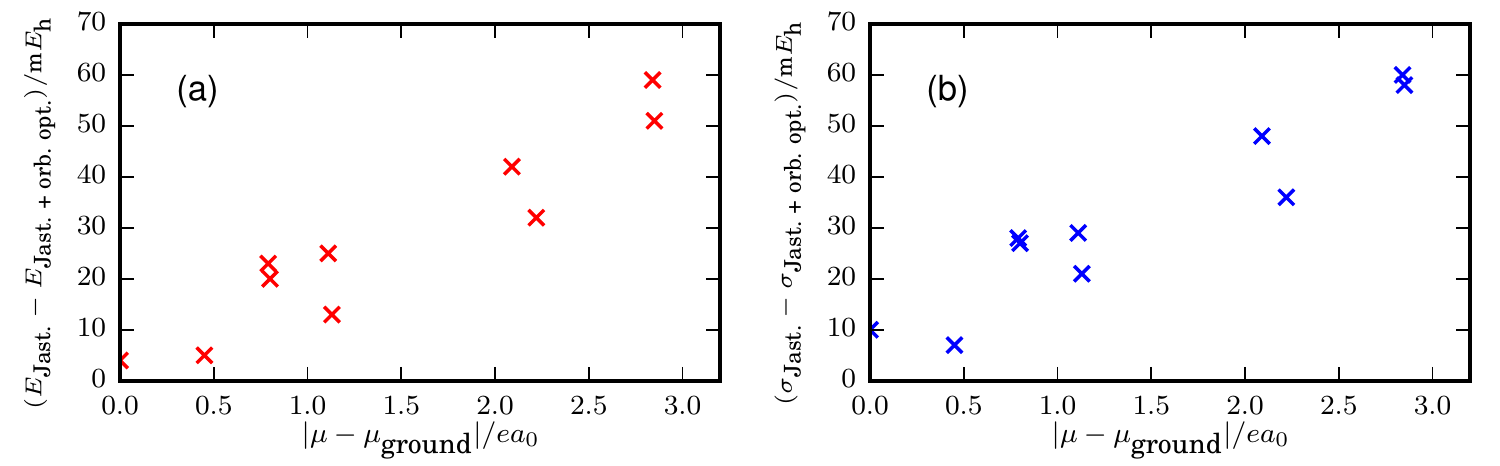}
\caption{Scatter plots demonstrating how energy and wave function improvement upon orbital optimization is correlated to dipole moment, relative to that of the ground state. The dipole moments presented are calculated from CIS solutions. (a) shows the difference in energy between two setups, one where the Jastrow alone is optimized, and another where the Jastrow and orbitals are optimized together. (b) demonstrates similar results, but for the difference in the standard deviation of the wave function energy. Clearly, the largest changes occur for excited states whose dipole moments differ the most from the ground state, with a strong correlation between the two. A large value of $|\mu - \mu_{\textrm{ground}}|$ corresponds to charge transfer.}
\label{fig:ch2o_scatter}
\end{figure*}
\begin{figure*}[t!]
\includegraphics{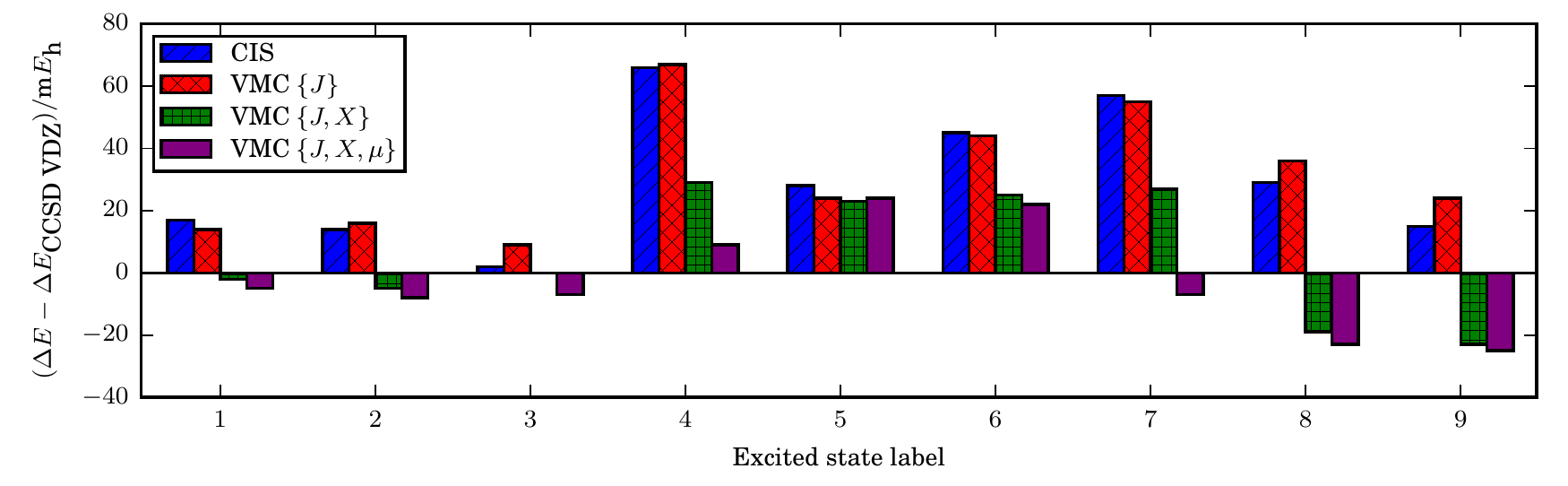}
\caption{Excitation energies (from the ground state) relative to those obtained from EOM-CCSD, for formaldehyde. Pseudopotentials\cite{Burkatzki2007} were used for C and O atoms, and the corresponding VDZ basis was used to construct determinants for the VMC calculations. Excited-state ordering is determined at the CIS level. CIS and EOM-CCSD results used the same pseudopotentials and basis sets. Because of the Jastrow factor, a direct comparison with VMC results is challenging, but general trends are visible. Optimization of the Jastrow alone $(\{ J \})$ makes little difference to excitation energies. Further optimization of orbitals $(\{ J, X \})$ leads to more significant changes, particularly for states 4, 7, 8, 9, which have significant CT character. The overshooting of states 8 and 9 could be related to basis set incompleteness error in EOM-CCSD benchmarks, which is particularly large for these two states. Excited state 5 has the smallest CT character of the excited states studied, and sees little improvement upon optimization, in line with the trend of Fig.~(\ref{fig:ch2o_scatter}).}
\label{fig:ch2o_bar}
\end{figure*}

The significance of CT character on optimization is now considered. At the RHF/CIS level the ground state has a dipole moment of $-1.134ea_0$ (away from the oxygen), while excited state dipole moments range from $-1.584ea_0$ to $+1.718ea_0$, with the latter being the state of most significant CT character. Fig.~(\ref{fig:ch2o_scatter}) shows how energy and the standard deviation of the energy change upon including orbitals in the wave function optimization. There is clearly a strong positive correlation between the CT character (as measured by the magnitude of the dipole moment relative to the ground state) and improvement upon orbital optimization. There is also clearly strong correspondence between energy reductions and standard deviation reductions, with near equality between the two in some cases. Energies and standard deviations are reduced by almost $2$eV in some cases.

Fig.~(\ref{fig:ch2o_bar}) shows excitation energies relative to EOM-CCSD, for CIS and VMC with various levels of wave function optimization. Care should be taken in the interpretation of these results. We note that both CIS and EOM-CCSD results are in the VDZ basis. Orbitals in the VMC calculation are also from the VDZ basis, but application of a Jastrow factor leads to a wave function that no longer exists in this basis, significantly reducing the variance of the wave function. However, the FDLR ansatz includes no dynamic correlation beyond the Jastrow, which mainly affects electron density near cusps, and is therefore mostly equal between valence-excited states. Dynamic correlation in the FDLR ansatz is further discussed in Section~\ref{sec:etfe}. Within the basis used, however, we expect that EOM-CCSD results to be accurate and a very significant improvement over CIS, and therefore sufficient for our current comparison.

VMC results are for: optimization of the Jastrow alone, $\{ J \}$; optimization of the Jastrow and orbitals together, $\{ J, X \}$; optimization of the Jastrow, orbitals and CIS coefficients together, $\{ J, X, \mu \}$. In the last case, the optimization is performed starting from the results of the Jastrow-orbital optimization, as discussed in Section~\ref{sec:LiF}.

In most cases, application of a Jastrow factor results in only small changes to excitation energies. Orbital optimization improves results more significantly. The states with strongest CT character are those with excited-state labels 4, 7, 8 and 9, where significant improvements are seen upon orbital optimization. In particular, states 8 and 9 have the largest values of $|\mu - \mu_{\textrm{ground}}|$, $\sim 3.0ea_0$ for both. These states overshoot the EOM-CCSD energies. This is understood by noting that these two EOM-CCSD excitation energies undergo a particularly large change upon moving to a valence-quadruple zeta (VQZ) basis set. Upon this change, EOM-CCSD excitations energies are lower than the presented VMC values. We also once again iterate that dynamic correlation could account for this quantitative difference. Most importantly, however, $\sigma$ for both states is reduced by $\sim 60$m$E_{\textrm{h}}$, demonstrating a significant improvement in the quality of the wave function in both cases.

\subsection{ETFE}
\label{sec:etfe}

For a demonstration in a somewhat larger system, ethylene-tetrafluoroethylene (ETFE) is considered, at distances (between the two molecules' center of masses (COM)) of $4$\AA \, and $8$\AA. As for other systems studied, pseudopotentials\cite{Burkatzki2007} and corresponding valence-double zeta basis sets were used for non-hydrogen atoms, and a standard cc-pVDZ basis set was used for hydrogen atoms, with 48 remaining electrons among 124 spatial orbitals (100 virtual). Thus, the total number of optimizable parameters is $2400$ in both $\{ X \}$ and $ \{ \mu \}$ sets, together with a smaller number of Jastrow parameters. This therefore proves a far sterner challenge, although the number of variables is small enough that the linear method eigenvalue problem may be solved exactly, without having to use more sophisticated schemes.\cite{Neuscamman2012,Zhao2017,Schwarz2017}

\begin{figure}[t!]
\includegraphics[width=60mm]{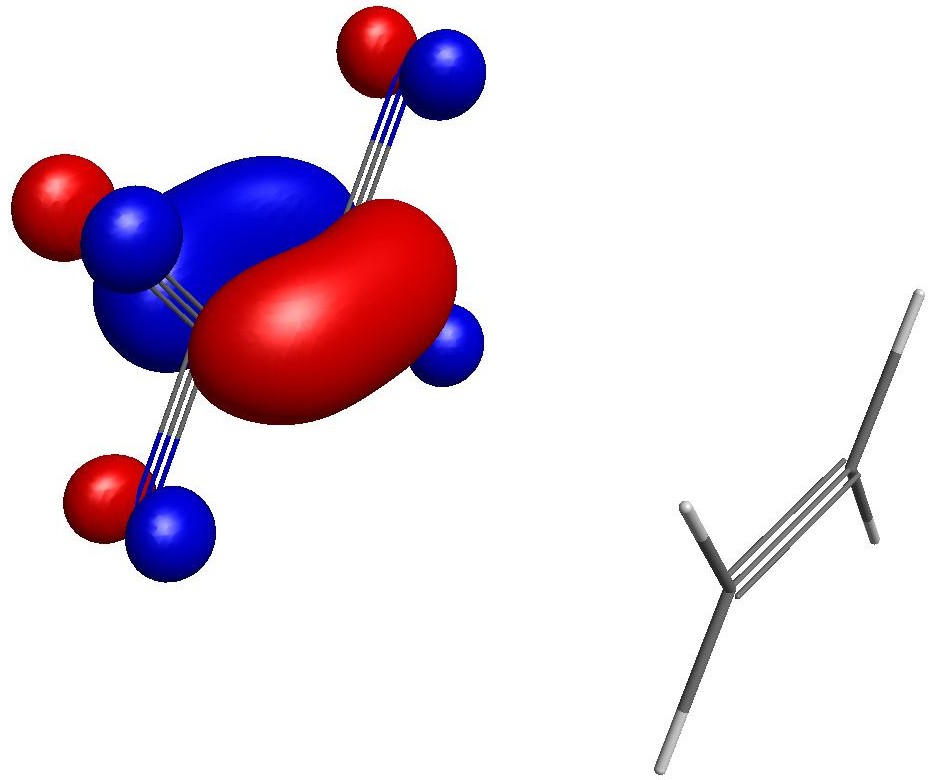}
\includegraphics[width=60mm]{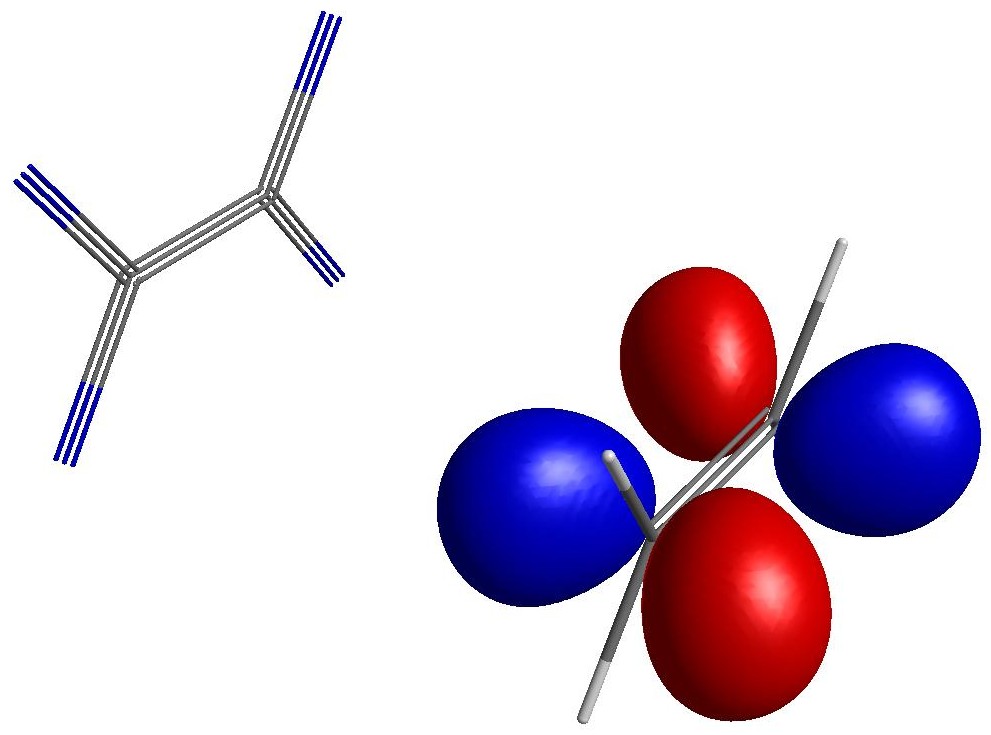}
\caption{Molecular orbitals of ETFE, at 4\AA \, COM separation. Top: HOMO of the tetrafluoroethylene molecule. Bottom: LUMO of the ethylene molecule. The CT state studied consists primarily of a single-electron excitation from the former to the latter orbital.}
\label{fig:etfe_orbs}
\end{figure}

ETFE has been important in the development of excited state methods: Dreuw \emph{et al.}\cite{Dreuw2003} used it to demonstrate clearly the failure of TDDFT in CT states. We investigate the same state from the study of Dreuw and co-workers, namely that corresponding to a transition from the HOMO of the tetrafluoroethylene to the LUMO of the ethylene. These orbitals are shown in Fig.~(\ref{fig:etfe_orbs}). For CIS and EOM-CCSD respectively, this transition has an amplitude of $0.775$ and $0.741$ at $4$\AA, and $0.99991$ and $0.954$ at 8\AA. At $4$\AA \, there are also significant non-CT determinants of CIS amplitudes $0.415$, $0.382$, and other important non-CT contributions. There is even a CIS amplitude of $0.132$ for a charge transfer determinant in the opposite direction: from the HOMO of the ethylene to the LUMO of the tetrafluoroethylene. Thus, both geometries clearly have CT character, although it is much stronger at $8$\AA.

Note that the studied CT state is actually the twelfth excited state (at the CIS level, at $4$\AA \, COM separation, and within the approximations applied - using a double-zeta basis set and BFD pseudopotentials). As such, it is a somewhat high-lying state. Nonetheless, we have been able to target it without apparent difficulty, and without having to calculate any lower-lying states. This is an important advantage of the state-specific approach used here.

We point out for completeness that the ethylene anion is not electronically stable.\cite{Simons1987} However, as a model system in a small basis, this does not affect the validity of our results or conclusions.

\begin{table}[t!]
\begin{center}
{\footnotesize
\begin{tabular}{@{\extracolsep{4pt}}lc@{}}
\hline
\hline
Number of samples & Energy/$E_{\textrm{h}}$ \\
\hline
$9.6 \times 10^4$   & -121.140(6) \\
$1.92 \times 10^5$  & -121.168(5) \\
$3.84 \times 10^5$  & -121.185(1) \\
$7.68 \times 10^5$  & -121.1917(6) \\
$1.536 \times 10^6$ & -121.1926(6) \\
$2.4 \times 10^6$   & -121.1944(3) \\
\hline
\hline
\end{tabular}
}
\caption{Convergence of the VMC energy (after simultaneous optimization of the Jastrow and orbitals) with number of samples (per linear method iteration), for the studied CT state of ETFE, and at a COM distance of 4\AA. All other simulation parameters were held constant across simulations (except the number of linear method itertations performed).}
\label{tab:etfe_error}
\end{center}
\end{table}

The solution of the linear method requires the solution of a stochastically-sampled eigenvalue problem. Since this is a non-linear problem, there will be a systematic bias in the solutions for a finite VMC sampling. For the LiF and formaldehyde cases this bias was not detectable compared to stochastic errors. However, in ETFE the bias becomes significant at a low number of samples. This is most likely due to the large number of parameters to be optimized, $\sim 2400$, and perhaps because of near-linear dependencies in orbital rotations. We note that optimization of the Jastrow alone does not show such an error, supporting these ideas. However, a similarly sized bias occurs for the ground state of ETFE, for which a single optimizable Slater-Jastrow function is used. Therefore, this bias does not appear to be larger for the FDLR ansatz specifically. In Table~\ref{tab:etfe_error}, convergence of this error is presented. It is seen that, although the bias is large with low numbers of samples, it converges rapidly, and these calculations typically used significantly less computational resources than large-scale QMC calculations. For example, convergence of the linear method with $1.536 \times 10^5$ samples (per linear method iteration) took around $24$ hours on $240$ CPU cores. The largest calculation considered for the following ETFE results was performed on NERSC's Edison machine, taking $\sim$35,000 core-hours for convergence.

\begin{table}[t!]
\begin{center}
{\footnotesize
\begin{tabular}{@{\extracolsep{4pt}}lcccc@{}}
\hline
\hline
& \multicolumn{4}{c}{Excitation energy, $\Delta E$/$E_{\textrm{h}}$} \\
\cline{2-5}
COM distance/\AA & CIS & EOM-CCSD & \{ J \} & \{ J, X \} \\
\hline
4.0 & 0.442 & 0.386 & 0.4385(6) & 0.4245(8) \\
8.0 & 0.513 & 0.456 & 0.5066(7) & 0.4650(8) \\
\hline
\hline
\end{tabular}
}
\caption{VMC excitation energies for ETFE at COM distances of $4$\AA \, and $8$\AA, compared to CIS and EOM-CCSD results. VMC results are shown for Jastrow-only optimization, $\{ J \}$, and simultaneous Jastrow-orbital optimization $\{ J, X \}$.}
\label{tab:etfe_results}
\end{center}
\end{table}

Table~\ref{tab:etfe_results} presents VMC excitation energies after optimization of the Jastrow alone, and optimization of the Jastrow and orbitals simultaneously. We do not include additional optimization of CIS coefficients, which seems to make no noticeable improvement within error bars. Furthermore, we do not repeat the optimization procedure with an updated value of the shift, $\omega$, as was done for LiF and formaldehyde, and discussed in Section~\ref{sec:var_principle}. In the latter cases it was found to make only small changes to excitation energies, and so for efficiency purposes the additional step was not performed.

The VMC excitation energies of Table~\ref{tab:etfe_results} are compared to CIS and EOM-CCSD results, performed in the VDZ basis, with pseudopotentials applied in all calculations. Optimizing the Jastrow alone leads to a relatively small decrease in the excitation energy of a few m$E_{\textrm{h}}$ at both geometries. At $8$\AA, optimization of the Jastrow and orbitals together reduces the CIS gap by around $48$m$E_{\textrm{h}}$, compared to $57$m$E_{\textrm{h}}$ with EOM-CCSD. At $4$\AA, this optimization reduces the energy by only $18$m$E_{\textrm{h}}$, compared to $56$m$E_{\textrm{h}}$ with EOM-CCSD. As noted, there is significantly greater CT character at $8$\AA, and so these results are in line with those of Fig.~(\ref{fig:ch2o_scatter}), demonstrating that orbital rotations have a larger impact in states with more significant CT nature.

Of course, this does not mean that non-CT states should be treated less accurately, but rather that the expected larger average error in CT states should be corrected by the optimization of orbitals, to bring accuracy of such states in line with non-CT ones. It is possible that the optimization at $4$\AA \, is not reaching the global minimum of $\Omega(\Psi)$, which is generally difficult to rule out in non-linear optimizations. However, we believe it more likely that this discrepancy is due to insufficient treatment of dynamic correlation. As discussed above, the state at $4$\AA \, COM separation has significant contributions from both CT and non-CT determinants. For the non-CT determinants, the RHF orbitals are mostly correct. For the CT determinant, the orbitals need significant relaxation. There are therefore competing requirements for the orbital optimization from the varying determinant contributions at $4$\AA, and the orbital optimization is likely to be less effective. This is essentially a case where correlation effects are very important: orbital relaxations for the remaining electrons depend on whether an electron is transferred.

The simple electron-electron and electron-nuclear Jastrow factors used here cannot capture this type of correlation. Such Jastrow factors are appropriate for treating short-range correlation, primarily due to cusp conditions. These effects are mostly the same between excitations that involve only valence electrons, as considered here. For very accurate excitation energies, an ansatz is required that can treat differential dynamic correlation between various excited states in a balanced manner. As one example of the inadequacy of the Jastrows used, we note that the electron-electron Jastrow factor is translationally invariant, and therefore does not depend on the position of a pair of electrons relative to nuclei. Clearly, we should expect electron-electron behavior to be dependent on distance to nuclear centers. More accurate treatments therefore include three-body (electron-electron-nuclear) Jastrow factors,\cite{Umrigar1988, Casula2004} and even four-body terms.\cite{Huang1997, Marchi2009, Needs2009} A further approach to accurate dynamic correlation is through backflow transformations.\cite{Lopez2006} Meanwhile, EOM-CC generally treats these various correlation effects in a more balanced and accurate manner.\cite{Krylov2008} This underlines the importance of accounting for dynamic correlation by a balanced approach. The FDLR ansatz and VMC optimization corrects the substantial orbital error in CT states, and we then expect that existing well-developed approaches, such as fixed-node DMC, will aid in improving the treatment of dynamic correlation. This was seen for LiF above, where the use of DMC led to substantially greater quantitative accuracy. Our current software does not allow us to make similar comparisons between DMC and CC for this particular system due to basis set mismatches, but we intend to pursue such comparisons in future work.

\section{Conclusion}
\label{sec:conclusion}

This article has introduced a simple and efficient wave function ansatz for use in real-space QMC methods, that encodes the full flexibility of CIS and orbital rotations in the difference of only two Slater determinant functions. This has been applied with a recently-introduced direct targeting approach for excited states, with the same basic $\mathcal{O}(N^4)$ scaling (or $\mathcal{O}(N^3)$ per sample) of traditional Slater-Jastrow QMC.

For CT states of LiF it was found that substantial improvements to the energy and wave function were obtained upon orbital relaxation. For a variety of states in formaldehyde, it was again found that the inclusion of orbital rotations is most crucial in CT states, with strong positive correlation between relative dipole moment and variance reduction. This was again observed in the ETFE system.

However, as is already understood, VMC for a basic Slater-Jastrow type function is not generally capable of capturing sufficient dynamic correlation for chemical accuracy, including the complex variations in dynamic correlation between different excited states, as required for accurate excitation energies. As such, the next step for this work is to treat dynamic correlation rigorously, either by use of an optimized FDLR wave function in DMC, or with more sophisticated Jastrow functions.\cite{Goetz2017, Neuscamman2013, Casula2004} Previous work by Filippi and coworkers has suggested that, with prior optimization of orbitals and expansion coefficients, accurate excited state results may be obtained via DMC.\cite{Schautz2004} We have performed preliminary DMC calculations on LiF, yielding high accuracy compared to EOM-CC in a very large basis set. This demonstrates that the finite-difference approach used here is indeed appropriate for use in diffusion Monte Carlo. With the low polynomial scaling of this approach, and the large-scale parallelism possible in QMC, we hope that this will lead to highly accurate excited-state results for CT states in systems containing several hundreds of electrons, as is already possible in ground state QMC.

\vspace{5mm}

\section{Acknowledgments}

We thank Ken Jordan for helpful discussion regarding this work. We gratefully acknowledge funding from the Office of Science, Office of Basic Energy Sciences, the US Department of Energy, Contract No. DE-AC02-05CH11231. Calculations on LiF and formaldehyde were performed on the Berkeley Research Computing Savio cluster. Larger calculations on ETFE were performed at the National Energy Research Scientific Computing Center, a DOE Office of Science User Facility supported by the Office of Science of the U.S. Department of Energy under Contract No. DE-AC02-05CH11231.


%

\end{document}